\documentclass[aps,twocolumn,a4paper,pra,superscriptaddress,floatfix,showpacs,nofootinbib]{revtex4}
\usepackage{amssymb}
\usepackage{amsmath}
\usepackage{dsfont}
\usepackage{amsfonts}
\usepackage{graphicx}
\usepackage{graphics}
\usepackage{color}

\begin{document}

\title{Casimir interaction between a dielectric nanosphere and a
metallic plane}
\author{Antoine Canaguier-Durand}
\author{Antoine G\'{e}rardin}
\author{Romain Gu\'{e}rout}
\affiliation{Laboratoire Kastler Brossel, CNRS, ENS, UPMC, case 74,
75252 Paris, France}
\author{Paulo A. Maia Neto}
\affiliation{Instituto de F\'{\i}sica, UFRJ, CP 68528, Rio de
Janeiro, RJ, 21941-909, Brazil}
\author{Valery V. Nesvizhevsky}
\affiliation{Institut Laue-Langevin, 6 rue Jules Horowitz, BP 156,
38042 Grenoble, France}
\author{Alexei Yu. Voronin}
\affiliation{P.N. Lebedev Physical Institute, 53 Leninsky prospect,
117924 Moscow, Russia}
\author{Astrid Lambrecht}
\author{Serge Reynaud}
\affiliation{Laboratoire Kastler Brossel, CNRS, ENS, UPMC, case 74,
75252 Paris, France}
\date{\today }

\begin{abstract}
We study the Casimir interaction between a dielectric nanosphere and
a metallic plane, using the multiple scattering theory. Exact
results are obtained with the dielectric described by a Sellmeier
model and the metal by a Drude model. Asymptotic forms are discussed
for small spheres, large or small distances. The well-known
Casimir-Polder formula is recovered at the limit of vanishingly
small spheres, while an expression better behaved at small distances
is found for any finite value of the radius. The exact results are
of particular interest for the study of quantum states of
nanospheres in the vicinity of surfaces.
\end{abstract}

\maketitle


\section{Introduction}

The Casimir effect, due to the scattering of quantum fluctuations of
the electromagnetic vacuum \cite{Casimir}, is the dominant
interaction between neutral bodies at distances large compared to
atomic scales \cite{Milonni,Lamoreaux} (more references in
\cite{Lambrecht10}). For this reason, it has a strong impact in
various important domains, such as atomic and molecular physics,
condensed matter and surface physics, chemical and biological
physics, micro- and nano-technology \cite{Parsegian}.

In this paper, we will consider the case of dielectric nanospheres
in the vicinity of a metallic surface. This study is part of a
discussion of the intriguing phenomenon of small heating of
ultra-cold neutrons (UCN) in traps
\cite{NesvizhevskyPAN02,LychaginPAN02,KartashovIJN07} which could be
explained by the interaction between UCN and nanospheres levitated
in the quantum states created by the interaction of nanospheres with
surfaces \cite{ILLLKB}. In order to characterize this phenomenon and
compute the properties of the quantum states, one needs to have a
detailed and careful treatment of the interaction potential. In
particular, as shown below, the commonly used Casimir-Polder formula
\cite{CasimirPolder} is not sufficient to this purpose.

Below we first recall how the scattering formula
\cite{LambrechtNJP06} can be applied to the study of a sphere of
radius $R$ at a distance $L$ of closest approach to the plane. We
then give numerical evaluations and graphical plots of the
interaction energy. The Casimir-Polder formula is recovered when the
radius of the nanosphere is smaller than all other length scales.
The short and long distance limits are then found to show subtle
interplays with the limit of small radius.

Important consequences of these results are obtained for the
behavior of the interaction at small distances. While the
Casimir-Polder energy scales as $L^{-3}$ at small values of $L$, the
full expression is found to be better behaved for any finite value
of $R$, which leads to a regular solution for the quantum states.


\section{Scattering formalism}

In this paper, we will not consider the effect of thermal
fluctuations on the interaction (they are expected to be small at
the not so large separations considered here). We thus start from
the scattering formula for the Casimir energy at zero temperature
\cite{LambrechtNJP06}
\begin{align}
& E=\hbar \int_{0}^{\infty }\frac{\mathrm{d}\xi}{2\pi} \ln \det
\mathcal{D}\,,\quad \mathcal{D} = \left( I-\mathcal{M}\right)
\label{scatterform} \\
& \mathcal{M} = \mathcal{R}_{S} e^{-\mathcal{K}\mathcal{L}}
\mathcal{R}_{P} e^{-\mathcal{K}\mathcal{L}} \,,\quad \mathcal{L}=L+R
\notag
\end{align}%
The Casimir energy is written in terms of reflection operators
$\mathcal{R}_{S}$ and $\mathcal{R}_{P}$ which describe the
diffraction by the sphere and the plate. These operators are
evaluated with reference points at the sphere center and at its
projection on the plane, respectively. The operator
$e^{-\mathcal{K}\mathcal{L}}$ accounts for one-way propagation along
the distance $\mathcal{L}=L+R$ separating these two points. The
operator $\mathcal{M}$ thus represents one round-trip propagation
inside the cavity formed by the two surfaces. All quantities are
written at imaginary frequencies $\omega =i\xi $ after a Wick
rotation.

The scattering formula (\ref{scatterform}) provides a compact way of
taking the multiple scatterings between the interacting bodies into
account. It can be considered as generalizing the
Dzyaloshinskii-Lifshitz-Pitaevskii formula \cite{DLP} to arbitrary
scattering properties of the two bodies. It can be applied in
various geometries and has in particular been recently used for
calculating the Casimir interaction between a metallic sphere and a
metallic plane
\cite{MaiaNetoPRA08,CanaguierPRL09,CanaguierPRL10,CanaguierPRA10}
(see also \cite{Emig08,Emig10}). In the following, we use the same
techniques and notations as in \cite{CanaguierPRL10,CanaguierPRA10}
and apply them to the case of a dielectric nanosphere above a
metallic plane.

The reflection on the plane is conveniently written by using a
plane-wave basis $|\mathbf{k},\pm ,p\rangle$ where $\mathbf{k}$ is
the wavevector component parallel to the plane $xy$ of the metallic
surface, $p=\mathrm{TE},\mathrm{TM}$ the polarization, $+/-$ the
upwards/downwards propagation direction. This basis is well adapted
to the description of the propagation operator
$e^{-\mathcal{K}\mathcal{L}},$ since $\mathcal{K}$ is thus diagonal
with elements $\kappa=\sqrt{\xi^2/c^2+k^2}$ representing the (Wick
rotated) wave-vector $z$-component for the imaginary frequency
$\xi$. The reflection operator $\mathcal{R}_{P}$ preserves all plane
wave quantum numbers but the propagation direction, and its elements
are given by the standard Fresnel specular reflection amplitudes for
an homogenous medium.

Then the reflection on the sphere is more easily written by using
the multipole basis $|\ell mP\rangle$, with $\ell(\ell+1)$ and $m$
denoting the angular momentum eigenvalues (with $\ell =1,2,...$,
$m=-\ell ,...,\ell $) and $P=E,M$ representing electric and magnetic
multipoles. The reflection operator $\mathcal{R}_{S}$ has its
elements given by the standard Mie scattering amplitudes. Thanks to
rotational symmetry, the operator $\mathcal{M}$ commutes with the
angular momentum operator $J_z$. Hence $\mathcal{M}$ is block
diagonal, and each block $\mathcal{M}^{(m)}$ (corresponding to a
given subspace $m$) yields an independent contribution to the
Casimir energy
\begin{eqnarray}
&&E=\frac\hbar\pi \int_0^\infty \mathrm{d}\xi ~\sum_m^\prime \ln
\det \left( I-\mathcal{M}^{(m)} \right)  \notag \\
&&\mathcal{M}^{(m)} = \left(
\begin{array}{cc}
\mathcal{M}_{EE}^{(m)} & \mathcal{M}_{EM}^{(m)} \\
\mathcal{M}_{ME}^{(m)} & \mathcal{M}_{MM}^{(m)}%
\end{array}%
\right) \label{blocks}
\end{eqnarray}%
The primed sum is a sum over positive integers with the term $m=0$
counted for its half. We have organized $\mathcal{M}^{(m)}$ in terms
of block-matrices built up on electric and magnetic contributions.
The corresponding matrix elements are given and discussed in
\cite{CanaguierPRA10}.

We now apply these formulas to the case of a dielectric nanosphere
above a metallic plane. In particular, the plots shown below will be
calculated for the case of interest for UCN studies \cite{ILLLKB},
namely a diamond nanosphere above a copper plane. We model copper
dielectric response with a Drude model
\begin{equation}
\varepsilon (i\xi )=1+\frac{\omega _{P}^{2}}{\xi (\xi +\gamma )}
\label{Drude}
\end{equation}%
This is written at imaginary frequencies $\omega =i\xi$, with
$\omega_P^2$ the squared plasma frequency proportional to the
density of conduction electrons in the metal, and $\gamma$ the
damping rate which measures the relaxation of these electrons. For
explicit calculations and plots, we will use the relations
$\omega_P=2\pi c/\lambda _{P}$ with the plasma wavelength
$\lambda_{P}=136$nm and $\gamma =0.0033\omega_{P}$. As $\gamma$ is
small when compared to $\omega_P$ for a good metal such as copper,
its influence is small at the not too large distances considered in
the present study (see \cite{Ingold09} for more details and
references). The diamond dielectric response is described by a
Sellmeier model
\begin{equation}
\varepsilon (i\xi )=1+\sum_{i}\frac{B_{i}\omega _{i}^{2}}{\omega
_{i}^{2}+\xi ^{2}}  \label{Sellmeier}
\end{equation}%
For diamond, a good enough description will be obtained with a
single component in this formula with $B_1=4.91$ and $\omega_1=2\pi
c/\lambda _1$ with the wavelength $\lambda _1=106$nm. The damping is
disregarded here since it do not play any significant role.

\begin{figure}[h]
\centering
\includegraphics[width=8cm]{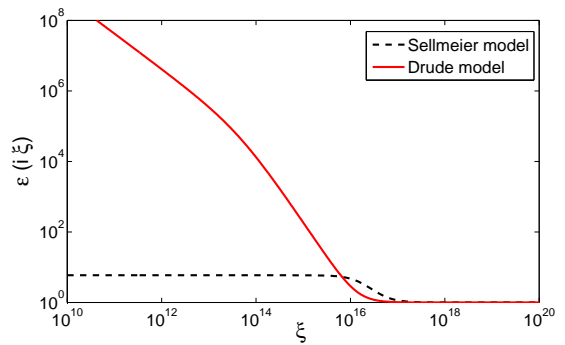}
\caption{Relative permittivity $\protect\varepsilon (i\protect\xi )$
for copper (Drude model; red line) and diamond (Sellmeier model;
black dashed).} \label{epsilon}
\end{figure}

The two permittivities are shown on Fig.\ref{epsilon}, where the
black dashed curve is for diamond, and the plain red curve for
copper. For copper, $\varepsilon(i\xi)$ is very large for $\xi$ much
smaller than $\omega_P $, which means that the metal tends to become
a very good reflector. For diamond, $\varepsilon(i\xi)$ tends to its
static value $\varepsilon(0)=1+B_1$ for $\xi$ much smaller than
$\omega_1$. For larger frequencies in contrast, the dielectric
properties become poorer for diamond as well as copper. As
$\omega_P$ and $\omega_1$ have similar values, we thus expect a
transition to take place between van der Waals and Casimir-Polder
regimes \cite{CasimirPolder} when the distance between two objects
is of the order of $\lambda_P$ or $\lambda_1$. The simple models
(\ref{Drude}-\ref{Sellmeier}) are sufficient for the purpose of the
present work. They could easily be improved to take into account
interband transitions for copper, multiple components and damping in
the Sellmeier model for diamond.


\section{Numerical evaluations}

For evaluating the determinant in (\ref{scatterform}-\ref{blocks}),
one needs to truncate the vector space at some maximum value
$\ell_{\mathrm{max}}$ of angular momentum
\cite{MaiaNetoPRA08,CanaguierPRL09,CanaguierPRL10,CanaguierPRA10}. A
qualitative understanding of the associated effects may be obtained
from the localization principle \cite{Nussenzveig}~: the value of
$\ell_{\mathrm{max}}$ required for a given accuracy level is
expected to scale with the size parameter $\hat{\xi}R$ (where
$\hat{\xi}=\xi/c$) which captures the dependence of scattering
amplitudes on the sphere radius. Meanwhile, the frequencies giving
the main contribution to the Casimir energy scale as $\hat{\xi} \sim
1/L$. As a consequence, the required $\ell _{\mathrm{max}}$ scales
as $R/L$ for intermediate and short separation distances. In the
present paper, we will be interested in nanospheres $R\leq 20$ nm so
that calculating with $\ell_{\mathrm{max}}=100$ will be sufficient
for a good accuracy for the results discussed below.

\begin{figure}[h]
\centering
\includegraphics[width=8cm]{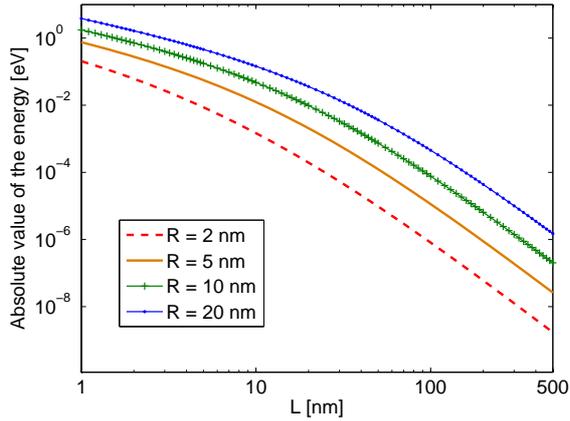}
\caption{Absolute value $\left\vert E\right\vert $ of the Casimir
energy $E$ (measured in eV) with respect to the distance $L$ for
nanospheres of radii $R=2,5,10,20$ nm. Logarithmic scales are used
on both axis. } \label{numerics}
\end{figure}

The numerical results are shown in Fig.\ref{numerics} for different
values of $R$ (2, 5, 10, and 20 nm). We plot the absolute value
$\left\vert E\right\vert$ of the Casimir energy ($E\leq 0$) with
respect to the distance $L$, from 1 to 500 nm. We see that two
regimes appear, which are reminiscent of non retarded van der Waals
regime at short distances \cite{Johannson}, and retarded
Casimir-Polder regime at large distances \cite{CasimirPolder}, with
different dependences upon the distance $L$ (see the slopes of the
curves) and the radius $R$ (see the vertical spacing between the
curves). Our results do not coincide with these well known limits
because they take into account higher order multipole contributions
to the scattering upon the sphere \cite{Noguez}. Of course, the
Casimir-Polder expression will be recovered for small values of the
radius, as shown in the next section.


\section{Limit of small nanospheres}

In this section, we consider the limit of a punctual sphere, when
the radius $R$ is smaller than any other length scale (in loose
terms, we take the limit $R\rightarrow 0$). We show that the
Casimir-Polder expressions are recovered, as expected. As the
derivations presented here are applied to dielectric nanospheres, we
will use the fact that the permittivity for the nanosphere remains
finite at all frequencies. We also keep the same description of the
dielectric response for the small values of the radius considered
here. Note that for metallic nanospheres in contrast, it would be
necessary to take into account the confinement effect for conduction
electrons \cite{Celep,Cottancin}.

We start from the formulas (\ref{scatterform}-\ref{blocks}) giving
the Casimir energy for a sphere of radius $R$ at a distance $L$ of
closest approach to a plane. Reflection on the plane is described by
Fresnel amplitudes $r_{\mathrm{TE}}$ and $r_{\mathrm{TM}}$ while
scattering on the sphere is described by Mie amplitudes $a_\ell$ and
$b_\ell$ (defined as in \cite{Bohren}). The following expressions,
valid at low values of the parameter $R\hat{\xi}$, are sufficient
for calculating the energy at the limit $R\rightarrow 0$
\begin{eqnarray}
a_\ell &\simeq&   (-1)^\ell~ \frac{\ell+1}{\ell \varepsilon +\ell+1}
\frac{(\varepsilon-1)\,\left(R\hat{\xi}\right)^{2\ell+1}}{(2\ell+1)!!
(2\ell-1)!!}
\label{Miell} \\
b_\ell  &\simeq&   (-1)^{(\ell+1)} ~ \frac{(\varepsilon-1)\,
\left(R\hat{\xi}\right)^{2\ell+3}}{(2\ell+3)!! (2\ell+1)!!} \notag
\end{eqnarray}%
As the dimensionless number $R\hat{\xi}$ is much smaller than unity
(for $L\hat{\xi} \sim 1$; see the discussion in the previous
section) and $\varepsilon$ remains finite at all frequencies, it
follows that all these amplitudes are small and, simultaneously that
the amplitude $a_{1}$ dominates all other Mie amplitudes.

The calculation of the energy is therefore much simpler than in the
general case. As a first simplification, one may indeed replace the
non linear expression (\ref{blocks}) by a linearized one
(perturbative approximation)
\begin{eqnarray}
&&E=-\frac\hbar\pi \int_0^\infty \mathrm{d}\xi ~\sum_m^\prime
\protect{\mathrm{tr}} \mathcal{M}^{(m)} \label{linblocks}
\end{eqnarray}%
Then, one may keep only the contributions to this sum which are
proportional to the amplitude $a_{1}$ (electric dipolar
approximation)
\begin{eqnarray}
&&E_{1}\simeq -\frac\hbar\pi \int_0^\infty \mathrm{d}\xi ~\left(
\frac{1}{2} \mathcal{M}_{EE}^{(0)}+\mathcal{M}_{EE}^{(1)}\right)  \\
&&\mathcal{M}_{EE}^{(0)} = -\frac{3}{2} \frac{a_{1}}{\hat{\xi}^{3}}
\int_0^\infty
~\frac{k^{3}\mathrm{d}k}{\kappa }r_{\mathrm{TM}}e^{-2\kappa L}  \notag \\
&&\mathcal{M}_{EE}^{(1)}=\frac{3}{4}\frac{a_{1}}{\hat{\xi}^{3}}
\int_0^\infty ~\frac{k\mathrm{d}k}{\kappa} \left(
\hat{\xi}^{2}r_{\mathrm{TE}}-\kappa ^{2}r_{\mathrm{TM}}\right)
e^{-2\kappa L}  \notag
\end{eqnarray}%
We also rewrite $a_{1}$ in terms of a dynamical electric
polarizability $\alpha(\xi)$ defined for the small nanosphere
($\alpha R^3$ is a reduced polarizability having the dimension of a
volume; the SI polarizability is $\varepsilon_0\alpha R^3$)
\begin{eqnarray}
&&a_{1} = -\frac{2}{3} \alpha R^3 \hat{\xi}^3 \quad,\quad \alpha
=\frac{\varepsilon-1}{\varepsilon+2}
\end{eqnarray}%
Collecting the results, we finally recover the full Casimir-Polder
formula \cite{CasimirPolder} as written in \cite{Dalvit08,Messina09}
\begin{eqnarray}
E_{1} & \simeq & -\frac{\hbar c R^3}{2\pi} \int_0^\infty
\mathrm{d}\hat{\xi}~\alpha (\hat{\xi})  \label{casimir_polder} \\
&&\hspace{-1cm}\times \int_0^\infty \frac{k\mathrm{d}k}{\kappa}
\left( \hat{\xi}^{2} |r_{\mathrm{TE}}| + \left( k^2+\kappa^2 \right)
|r_{\mathrm{TM}}|\right) e^{-2\kappa L}  \notag
\end{eqnarray}%
We have used the fact that $r_{\mathrm{TE}}\left( i\xi \right) <0$ and $r_{%
\mathrm{TM}}\left( i\xi \right) >0$.

We repeat at this point that the formula (\ref{casimir_polder}) has
been obtained after two simplifications corresponding to the
perturbative approximation and electric dipolar approximation. As
the Casimir-Polder interaction between atoms \cite{CasimirPolder},
it contains in particular the limits of non retarded van der Waals
and retarded Casimir-Polder expressions \cite{Dalvit08,Messina09}.
The energy scales in both cases as $R^{3}$ that is also the volume
of the sphere. This result means that the nanosphere behaves at the
limit $R\rightarrow 0$ as a big atom with an electric polarizability
$\alpha R^{3}$. As we will see in the next section, this simple
behavior does not remain true for arbitrary values of the radius
$R$.


\section{Asymptotic behavior at short and long distances}

In this section, we discuss the asymptotic behaviors of the Casimir
energy $E$ at short and long distances.

As a first step to this aim, we write a Casimir-Polder formula
$E_{\mathrm{CP}}$ deduced from $E_{1}$ in the limit $L\gg
\lambda_{P},\lambda _{1}$ where copper may be considered as
perfectly reflecting and diamond as having a constant electric
polarizability
\begin{eqnarray}
&&E_{\mathrm{CP}}=-\frac{4\pi c_{4}R^{3}}{3L^{4}}    \label{ECP} \\
&&c_{4}=\frac{ 9\hbar c\alpha _{0}}{32\pi ^{2}} \,,\quad\alpha
_{0}=\frac{\varepsilon (0)-1}{\varepsilon (0)+2} \notag
\end{eqnarray}%
We proceed similarly with the van der Waals prediction
$E_{\mathrm{vdW}} $ deduced from $E_{1}$ for $L\ll \lambda
_{P},\lambda _{1}$
\begin{eqnarray}
&&E_{\mathrm{vdW}}=-\frac{4\pi c_{3}R^{3}}{3L^{3}}  \label{EvdW} \\
&&c_{3}=\frac{3\hbar c\alpha _{0}}{16\left( \sqrt{2}\lambda _{P}+\sqrt{%
1-\alpha _{0}}\lambda _{1}\right) }  \notag
\end{eqnarray}
The values $E_{\mathrm{CP}}$ and $E_{\mathrm{vdW}}$ are equal at a
crossing length
\begin{equation}
L_\ast = \frac{c_4}{c_3} = \frac{3 \left( \sqrt{2}\lambda
_{P}+\sqrt{1-\alpha _{0}}\lambda _{1}\right)}{2\pi^2}
\end{equation}%
that is approximately $39$nm with the values corresponding to copper
and diamond.

As expected from already presented qualitative arguments, the exact
Casimir expression $E$ (given by eq.\ref{scatterform}) is well
approximated by $E_{\mathrm{CP}}$ when $R$ is the smallest length
scale and $L$ the largest one
\begin{equation}
E\simeq E_{\mathrm{CP}} \,,\quad R \ll L_\ast \ll L \label{condCP}
\end{equation}%
Meanwhile, the exact Casimir expression $E$ is well approximated by
$E_{\mathrm{vdW}}$ when $R$ is the smallest length scale and $L$
smaller than $L_\ast$
\begin{equation}
E\simeq E_{\mathrm{vdW}} \,,\quad R \ll L \ll L_\ast \label{condvdW}
\end{equation}
Of course, there exist a variety of behaviors when the two latter
conditions are not met.

\begin{figure}[h]
\centering
\includegraphics[width=9cm]{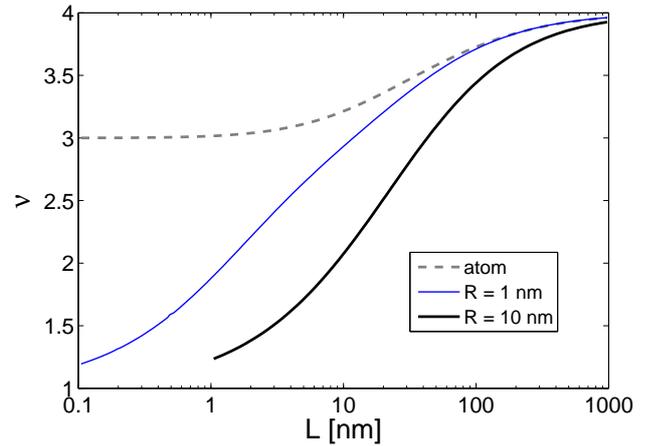} %
\caption{Logarithmic slope $\protect\nu $ as a function of the
distance $L$. Plain curves represent the nanosphere case with
different values for the radius $R$, dashed curve represents the
atomic limit, that is a nanosphere with $R\rightarrow 0$. }
\label{powerL}
\end{figure}

In order to explore this variety, we plot on Fig.\ref{powerL} the
logarithmic slope (log-log derivative) of the energy $|E|$ versus
distance $L$
\begin{equation}
\nu =-\frac{\partial \ln |E|}{\partial \ln L}=\frac{LF(L)}{E(L)}\quad ,\quad
F(L)=-\frac{\partial E}{\partial L}  \label{slopeL}
\end{equation}%
The parameter $\nu $ would be a constant if the energy $|E|$ obeyed
a power law dependence $1/L^{\nu }$ (for example $\nu =4$ for
$E_{\mathrm{CP}}$\ or $\nu =3$ for $E_{\mathrm{vdW}}$). In the
general case, $\nu $ depends on $L$ and can thus be understood as
describing a `local' power law in the vicinity of $L$.

We see on Fig. \ref{powerL} that $\nu$ tends to the expected value 4
at large distances. At small distances in contrast, the van der
Waals value $\nu =3$ is never a good approximation, which can be
understood by inspecting the conditions for (\ref{condvdW}) to be
true. For any finite value of the radius $R$, we have indeed to
cross the conditions $L\sim R$ when the distance is decreased and
(\ref{condvdW}) can no longer be valid after this crossing.

\begin{figure}[h]
\centering
\includegraphics[width=9cm]{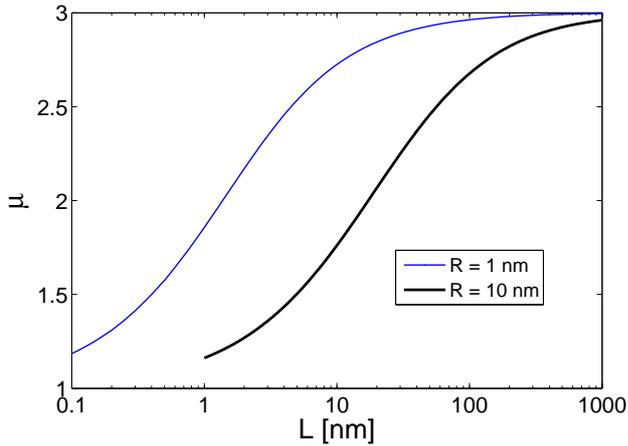}
\caption{Logarithmic slope $\protect\mu $ for variation with the
radius $R$, as a function of the distance $L$. Plain curves
represent the nanosphere case with different values for the radius
$R$, the atom case being in comparison a constant $\protect\mu =3$.
} \label{powerR}
\end{figure}

Another facet of the same problem becomes apparent when looking at
the dependence of $E$ versus $R$. We plot on Fig.\ref{powerR} a
logarithmic slope $\mu$ calculated as in (\ref{slopeL})
\begin{equation}
\mu = \frac{\partial \ln |E|}{\partial \ln R} \label{slopeR}
\end{equation}%
The parameter $\mu$ would be a constant if $|E|$ obeyed a simple
power law dependence $R^\mu$. In particular the volumetric value
$\mu =3$ is obtained for both $E_\mathrm{CP}$\ and $E_\mathrm{vdW}$.
We see on Fig. \ref{powerR} that $\mu$ approaches this value at the
limit of large distances, but departs from it everywhere else, anew
indicating that $E_\mathrm{vdW}$ is not a good approximation.

We now give an improved version of the van der Waals formula
(\ref{EvdW}) which explains some of the features of the exact energy
$E$. As $E_{\mathrm{vdW}}$ has been demonstrated above in the limit
of a punctual sphere $R\rightarrow 0$, we may improve it for a
finite size of the sphere through a pairwise summation over the
volume. We thus obtain the Hamaker expression \cite{Hamaker}
\begin{eqnarray}
\overline{E_{\mathrm{vdW}}} &\simeq &-\pi c_{3}\left( \frac{2R\left(
L+R\right) }{L\left( L+2R\right) }-\ln \frac{L+2R}{L}\right) \label{Ham} \\
\overline{E_{\mathrm{CP}}} &\simeq &-\frac{4\pi
c_{4}R^{3}}{3L^{2}\left( L+2R\right) ^{2}}  \notag
\end{eqnarray}
For completeness, we did proceed similarly with the Casimir-Polder
formula $E_{\mathrm{CP}}$.

These results allow one to understand the behaviors apparent on
Figs.\ref{powerL}-\ref{powerR}. Let us again consider that we start
from small nanospheres $R\ll L_\ast$ at large distances $L\gg
L_\ast$. Using the expression $E_\mathrm{CP}$, one obtains $\nu=4$
and $\mu=3$. When the distance $L$ is decreased, we cross two
transitions $L\sim L_\ast$ and $L\sim R$ and end up with the formula
$\overline{E_\mathrm{vdW}}$ for which we get $\nu=1$ and $\mu=1$.
This line of reasoning reproduces the global variations seen on
Figs.\ref{powerL}-\ref{powerR}.

\begin{figure}[h]
\centering
\includegraphics[width=9cm]{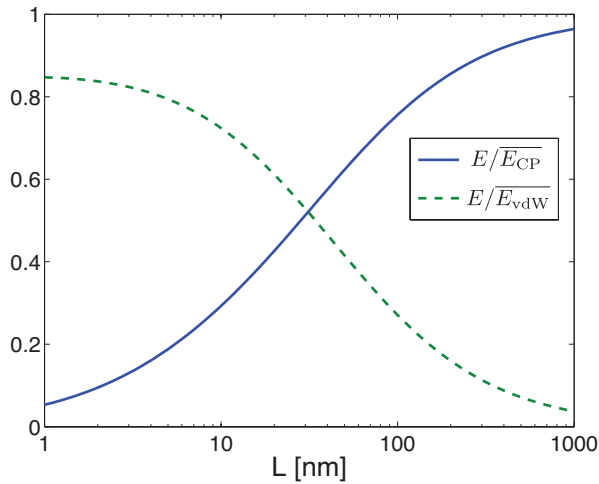}
\caption{Ratios of the exact energy $E$ to the expressions
$\overline{E_{\mathrm{CP}}}$ (plain blue curve) and
$\overline{E_{\mathrm{vdW}}}$ (dashed green curve). The two curves
are calculated for $R=$10nm.} \label{energy_comparisons}
\end{figure}

In order to assess the quality of the estimations (\ref{Ham}), we
now plot on Fig.\ref{energy_comparisons} the ratios
$E/\overline{E_\mathrm{CP}}$ and $E/\overline{E_\mathrm{vdW}}$. As
expected, we find that $\overline{E_\mathrm{CP}}$ tends to reproduce
the result $E$ of the full numerical computation at large distances.
We also see that $\overline{E_\mathrm{vdW}}$ obeys the same power
law than $E$ at small distances (ratio tending to a constant value),
but fails to predict the correct magnitude (the limit of the ratio
is not 1). This feature can be understood through a close inspection
of the case where $L$ is the smallest of all length scales.

When $L\ll R$, we can use the proximity force approximation and
express the plane-sphere result in terms of the plane-plane one
\cite{Derjaguin}. We are thus left with the evaluation of the
Casimir effect between copper and diamond plane plates. As $L\ll
L_\ast$, it is then possible to use the method designed in
\cite{Genet} to find
\begin{eqnarray}
&&E_\mathrm{PFA}\simeq -\pi c^\prime_3 \frac RL \,,\quad L \ll R
\,,\, L_\ast \label{PFA}
\end{eqnarray}
This expression shows the same functional dependencies than the
Hamaker expression $\overline{E_\mathrm{vdW}}\sim -\pi c_{3} R/L$
while giving a different proportionality constant. This difference
is due to the fact that (\ref{Ham}) has been obtained through a
pairwise integration of van der Waals forces whereas (\ref{PFA}) has
been calculated by taking into account the multiple interferences
occuring in the Fabry-Perot cavity \cite{Genet}. The numerical
values thus obtained ($c^\prime_3=0.84\,c_3$ with the numbers
corresponding to diamond and copper) explain the behavior seen on
Fig.\ref{energy_comparisons}.


\section{Conclusions}

In this paper, we have presented an exact calculation of the Casimir
interaction between a dielectric nanosphere and a metallic plane,
using the multiple scattering formalism as developed recently for
the plane-sphere geometry
\cite{MaiaNetoPRA08,CanaguierPRL09,CanaguierPRL10,CanaguierPRA10}.
In order to discuss qualitatively the results obtained in this
manner, we have also investigated the limits of a punctual sphere as
well as the asymptotic behaviors at short and long distances.

This study has important applications for discussing the intriguing
phenomenon of heating of ultra-cold neutrons in traps
\cite{NesvizhevskyPAN02,LychaginPAN02,KartashovIJN07}. This heating
could be explained by the interaction between UCN and nanospheres
levitated in the quantum states created by their interaction with
surfaces \cite{ILLLKB}. In order to characterize quantitatively this
phenomenon, a detailed knowledge of the interaction potential is
required. In particular, the small-distance behavior of the Casimir
energy plays a crucial role in the determination of the quantum
states obtained by solving the Schr\"{o}dinger equation for the
wavefunction of the nanosphere in this potential.

The commonly used Casimir-Polder formula, which also corresponds to
the limit of our calculations for a vanishingly small radius $R$,
leads to significant difficulties since it predicts a power law $|E|
\propto R^3/L^3$ in the vicinity of the surface and thus leads to a
ill-behaved Schr\"{o}dinger problem. The exact solution presented in
the present paper for a finite value of the radius $R$ predicts a
smoother power law $|E| \propto R/L$ in the vicinity of the surface
and thus leads to a regular solution for the Schr\"{o}dinger
equation \cite{ILLLKB}.

The plate roughness has been disregarded here and it is treated in a
phenomenological manner in \cite{ILLLKB}. It would be interesting to
analyze its effect by using techniques already developed for
treating the scattering on rough surfaces
\cite{MaiaNetoEPL05,MaiaNetoPRA05,Contreras}.

It would also be worth investigating the same problem for the
interaction between an atom and a plane. In analogy with the
discussion of the present paper, taking into account the higher
order multipoles and multiple interferences could lead to an
expression of the energy more regular than with the commonly used
electric dipole approximation.

\acknowledgments The authors thank M.-C. Angonin, I. Cavero-Pelaez,
G.-L. Ingold, R. Messina, H.M. Nussenzveig, S. Pelisson and P. Wolf
for discussions, and the ESF Research Networking Programme CASIMIR
(www.casimir-network.com) for providing excellent opportunities for
discussions on the Casimir effect and related topics. PAMN thanks
CNPq and FAPERJ for financial support.

\end{document}